\documentclass[fleqn,usenatbib,twocolumn]{mnras}
\usepackage{pdflscape}
\usepackage{newtxtext,newtxmath}
\usepackage{multirow}
\usepackage[T1]{fontenc}

\usepackage{comment}
\usepackage{enumerate}

\DeclareRobustCommand{\VAN}[3]{#2}
\let\VANthebibliography\thebibliography
\def\thebibliography{\DeclareRobustCommand{\VAN}[3]{##3}\VANthebibliography}

\usepackage{graphicx}	% Including figure files
\usepackage{amsmath}	% Advanced maths commands
\title[Quantifying the randomness and scale invariance of repeating FRBs]{Quantifying the randomness and scale invariance of the repeating fast radio bursts}

\author[Y. Sang and H.-N. Lin]{
Yu Sang$^{1}$,
Hai-Nan Lin$^{2, 3}$\thanks{Corresponding author: linhn@cqu.edu.cn}\\
$^{1}$Center for Gravitation and Cosmology, College of Physical Science and Technology, Yangzhou University, Yangzhou 225009, China\\
$^{2}$Department of Physics, Chongqing University, Chongqing 401331, China\\
$^{3}$Chongqing Key Laboratory for Strongly Coupled Physics, Chongqing University, Chongqing 401331, China
}

\date{Accepted XXX. Received YYY; in original form ZZZ}

\pubyear{2024}

\begin{document}
\label{firstpage}
\pagerange{\pageref{firstpage}--\pageref{lastpage}}
\maketitle

\begin{abstract}
The statistical properties of energy and waiting time carry essential information about the source of repeating fast radio bursts (FRBs). In this paper, we investigate the randomness of energy and waiting time using four data samples from three extremely active repeating FRBs observed by the Five-hundred-meter Aperture Spherical radio Telescope (FAST). We report the deviation from complete randomness of the burst activity using three statistics, i.e., Hurst exponent, Pincus index and non-Gaussian probability density distribution of fluctuations. First, the Hurst exponent greater than 0.5 reveals that there is long-term memory in the time series of energy and waiting time. Second, the deviation of the Pincus index from 1.0 manifests that the time series is not completely random. Finally, the fluctuations of energy and waiting time follow the scale-invariant $q$-Gaussian distribution. All these statistical properties imply that, although the time series of repeating FRBs seems to be irregular, they are not completely random, similar to the features of self-organized criticality.
\end{abstract}


%%%%%%%%%%%%%%%%%%%%%%%%%%%%%%%%%%%%%%%%%%%%%%%%%%

%%%%%%%%%%%%%%%%% BODY OF PAPER %%%%%%%%%%%%%%%%%%

\section{Introduction}\label{sec:intro}

Fast radio bursts (FRBs) are bright astronomical transients with milliseconds duration in radio bands \citep{Lorimer:2007qn,Cordes:2019cmq,Petroff:2019tty,Platts:2018hiy,Zhang:2020qgp,Xiao:2021omr,Zhang:2022uzl}. 
The large dispersion measure implies that they originate from the extragalactic Universe rather than the Milky Way, which has been proven by the direct localization of the host galaxy and the measurement of redshift \citep{Keane:2016yyk,Chatterjee:2017dqg,Tendulkar:2017vuq}. Since the first discovery in 2007 \citep{Lorimer:2007qn}, hundreds of FRB sources have been observed by different telescopes \citep{Petroff:2016tcr,CHIMEFRB:2021srp,Xu:2023did}. Although most FRBs (more than 90 percent) seem to be one-off events, a plenty of them are verified to be repeaters. For example, in the first CHIME/FRB catalog \citep{CHIMEFRB:2021srp}, 18 FRBs are repeaters and 474 FRBs are appreantly one-off events. The first and best-known repeater is FRB 20121102A, which keeps active in a long period spanning several years \citep{Spitler:2016dmz,Chatterjee:2017dqg,Michilli:2018zec,Gajjar:2018bth,Zhang:2018jux,Gourdji:2019lht,Li:2021hpl}. The physical mechanism of FRBs is still under extensive debate, and many theoretical models are proposed to explain the origin of FRBs, see e.g. \citet{Platts:2018hiy} for the review of FRB models.

The statistical study of energy and waiting time helps to reveal the mystery of radiation mechanism of FRBs, because these observed parameters carry information about the source activity. Actually, the statistical properties of energy and waiting time of FRBs have been extensively studied in literature \citep{Lu:2016fgg,Li:2016qbl,Wang:2016lhy,Wang:2017agh,Macquart:2018jlq,Opperman:2017kql,Wang:2019sio,Lu:2019pdn,Lin:2019ldn,Wang:2019suh,Zhang:2021ztz,Wang:2022gmu,Sang:2023zho}. The burst energy of repeating FRBs is usually assumed to follow the power law distribution. For example, \citet{Wang:2016lhy} found that the cumulative distributions of peak flux and fluence of 17 bursts from repeating FRB 20121102A show the power law form. Using six samples from different telescopes at different frequencies, \citet{Wang:2019sio} found that the power law indices are in a narrow range, indicating a universal energy distribution for FRB 20121102A. The waiting time of repeating FRBs can be fitted by several distributions, e.g. the power-law distribution \citep{Wang:2016lhy}, the exponential distribution \citep{Wang:2017agh}, or the Weibull distribution \citep{Opperman:2017kql,Zhang:2021ztz}. In addition, \citet{Lin:2019ldn} showed that the distribution of energy and waiting time of FRB 20121102A can be well fitted by the bent power-law, while the fluctuations of energy and waiting time follow the scale-invariant $q$-Gauss distribution.

The operation of the Five-hundred-meter Aperture Spherical radio Telescope (FAST) significantly enlarges the repeating FRB samples, thus allows us to statistically study the repeating FRBs in details. For example, \citet{Li:2021hpl} reported the detection of 1652 bursts from the repeating FRB 20121102A, and found that the burst energy can be well fitted by a lognormal distribution plus a generalized Cauchy function, and the waiting time is consistent with the superposition of two lognormal distributions. \citet{Xu:2021qdn} found that the energy of 1863 bursts from the repeating FRB 20201124A can be fitted by a broken power law function, while the waiting time can be fitted by the superposition of three lognormal distributions. \citet{Zhang:2023eui} also found that the energy of 1076 burst from FRB 20220912A can be well described using a broken power-law function, and the waiting time can be described using two lognormal distributions. \citet{Sang:2023zho} found that the energy and waiting time of FRB 20121102A and FRB 20201124A can be well fitted by the bent power law, and the power-law index keeps steady over time. \citet{Zhang:2022rib} found that the distribution of burst energy from FRB 20201124A can be well modelled by an exponentially connected broken-power law function. \citet{Zhang:2023fmn} found that repeating FRBs show the feature of Brownian motion
on the time-energy bivariate space. \citet{Wang:2023wcb} found that repeating FRBs have long-term memory over a large
range of timescales, from a few minutes to about an hour, see also \citet{Wang:2023sjs}.

In this paper, we further study the statistical properties of repeating FRBs based on three statistics, i.e. the Hurst exponent, the Pincus index, and the fluctuations of time series. Four data samples from three extremely active repeating FRBs detected by the FAST telescope are considered, namely FRB 20121102A, FRB 20201124A and FRB 20220912A. The first sample is FRB 20121102A, which includes 1652 bursts detected in a total of 59.5 observation hours spanning two months in 2019 \citep{Li:2021hpl}. The host of FRB 20121102A is well identified at redshift $z=0.193$, which corresponds to a luminosity distance $D_L=949$ Mpc \citep{Tendulkar:2017vuq}. From April 1 to June 11 in 2021, the FAST telescope recorded 1863 bursts (hereafter FRB 20201124A-I) from the repeating FRB 20201124A during a total of 82 hours observation time \citep{Xu:2021qdn}. Three months later, 881 additional bursts (hereafter FRB 20201124A-II) were detected by FAST at the end of September in 2021, in a total of 19 hours observation time \citep{Zhang:2022rib}. Considering the possible temporally evolving burst activity, we study these two samples from the same FRB source separately. The redshift of FRB 20201124A is $z=0.09795$, corresponding to a luminosity distance $D_L=453.3$ Mpc \citep{Xu:2021qdn}. The FAST telescope recorded 1076 bursts from FRB 20220912A in a total of 8.67 observation time in 2022 \citep{Zhang:2023eui}. The host galaxy of FRB 20220912A was localized at redshift $z=0.0771$, corresponding to a luminosity distance $D_L=360.86$ Mpc \citep{Ravi:2022rbq}.

Another sample FRB 20190520B was studied in a similar way in two previous works \citep{Zhang:2023fmn,Yamasaki:2023fud}. The observed bursts of FRB 20190520B are much less than the four samples used in our paper. For example, the numbers of published bursts are 79 from FAST \citep{Niu:2021bnl} and 113 from Parkes \citep{Anna-Thomas:2022yvr}. Compared with other four samples, FRB 20190520B is not statistically significant. Moreover, the works of \citet{Zhang:2023fmn} and \citet{Yamasaki:2023fud} combined the data of FRB 20190520B observed by two different telescopes (FAST and Parkes). Since different telescopes have different sensitivity and selection effects, the combination of different data samples may cause bias on the statistical results. Therefore, we do not consider FRB 20190520B here.

Based on the public datasets, we calculate the waiting time and energy of each burst. The waiting time is defined as the observed  time interval between two successive bursts, $\delta t = t_{i+1} - t_i$, where $t_i$ is the arrival time of the $i$-th burst.
Since the observation of FAST is discontinuous, with 1 or 2 hours in each day, we pick the waiting times smaller than 1 hour to discard the long observation gaps between different observation sessions. 
After the selection, the number of waiting time for each sample is as follows: 1612 in FRB 20121102A, 1815 in FRB 20201124A-I, 877 in FRB 20201124A-II, and 1058 in FRB 20220912A.
The burst energy is calculated by \citep{Zhang:2023eui} 
\begin{equation}\label{eq:energy}
    E = \frac{4\pi}{1+z} \left(\frac{D_\mathrm{L}}{10^{28}\mathrm{cm}}\right)^{2} \left(\frac{F}{\mathrm{Jy}\cdot \mathrm{ms}}\right)\left(\frac{\Delta\nu}{\mathrm{GHz} }\right) 10^{39} \mathrm{erg},
\end{equation}
where $D_L$ is the luminosity distance, $F$ is the fluence, and $\Delta \nu=500\,\rm MHz$ is the observation bandwidth. 
For the energy data, we utilize the full datasets without excluding any data points. The number of energy in each sample is 1652 in FRB 20121102A, 1863 in FRB 20201124A-I, 881 in FRB 20201124A-II, and 1076 in FRB 20220912A.

The rest parts of this paper are arranged as follows: In Section \ref{sec:hurst}, we calculate the Hurst exponent and investigate the long-term memory of FRBs. In Section \ref{sec:pincus}, we calculate the Pincus index and investigate the randomness of FRBs. The fluctuations of burst energy and waiting time are investigated in Section \ref{sec:fluctuation}. Finally, discussion and conclusions are presented in Section \ref{sec:conclusion}.

\section{Memory and Hurst Exponent}\label{sec:hurst}

In this section, we study the long-term memory of waiting time and energy of repeating FRBs. The long-term memory of a time series can be quantified by the Hurst exponent $H$ \citep{hurst1956problem,hurst1957suggested}. $H = 0.5$ means no long-range correlations in the data,  $H < 0.5$ means negative long-range correlations, and $H > 0.5$ means positive long-range correlations.

The Hurst exponent of a time series, e.g., the waiting time or energy of a repeating FRB, is calculated using the rescaled range analysis method \citep{Weron_2002,MERAZ2022126631}. A time series of length $N$ is divided into $l$ subseries of length $n$ with no overlap. For each of the subseries $X_m= \{X_{1,m}, X_{2,m}, ..., X_{n,m}\}$, with $m=1,2,...,l$, the Hurst exponent is calculated according to the following setps: 
\begin{enumerate}[(1)]
\item  Compute the mean value $E_m$ and the standard deviation $S_m$; 
\item  Compute the mean-adjusted time series $Y_{i,m}=X_{i,m}-E_m$ for $i=1,...,n$; 
\item  Calculate the cumulative deviation series $Z_{i,m}=\sum_{j=1}^i Y_{j,m}$ for $i=1,...,n$;
\item  Calculate the series range $R_m = \max \{Z_{1,m},...,Z_{n,m}\} - \min \{Z_{1,m},...,Z_{n,m}\}$;
\item  Calculate the rescaled range $R_m/S_m$;
\item  Repeat steps (1) -- (5) to obtain the rescaled range for all subseries, and calculate the mean value of the rescaled range for all subseries of length $n$: 
$$
(R/S)_n  = \frac{1}{l}\sum_{m=1}^l R_m/S_m.
$$
\item Change the subseries length $n$ and construct a series of the rescaled range;
\item The rescaled range series $(R/S)_n$ for a self-similar process follows a power-law function
$$
(R/S)_n = Cn^H.
$$
Here $H$ is the Hurst exponent, which can be estimated by a simple linear regression over 
$$
\ln (R/S)_n = \ln C + H \ln n.
$$
\end{enumerate}

\begin{figure*}
    \centering
	\includegraphics[width=0.48\textwidth]{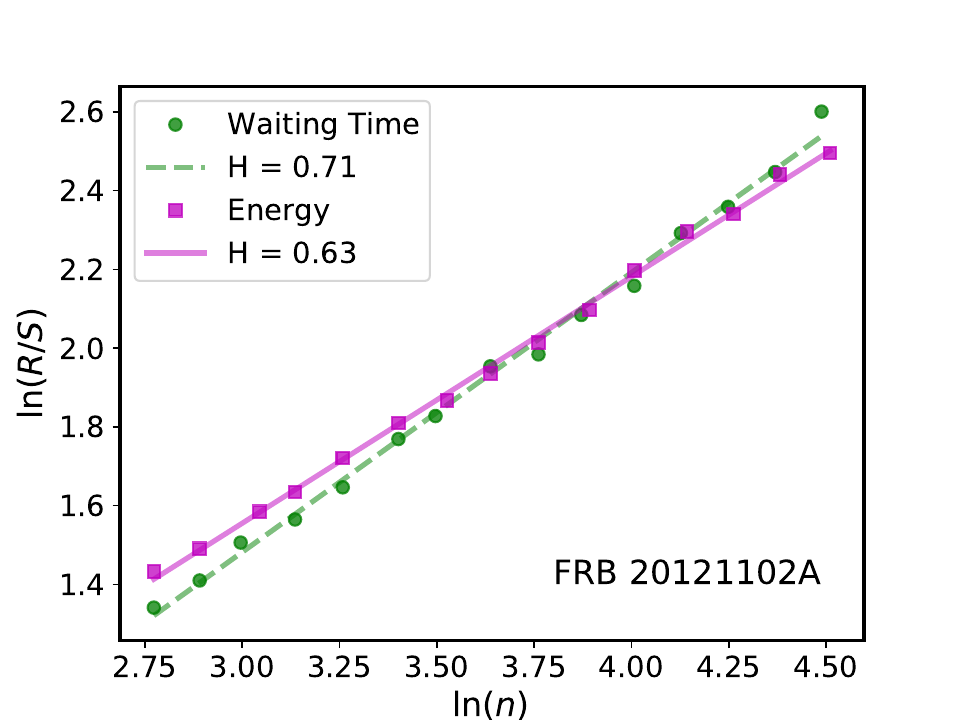}
	\includegraphics[width=0.48\textwidth]{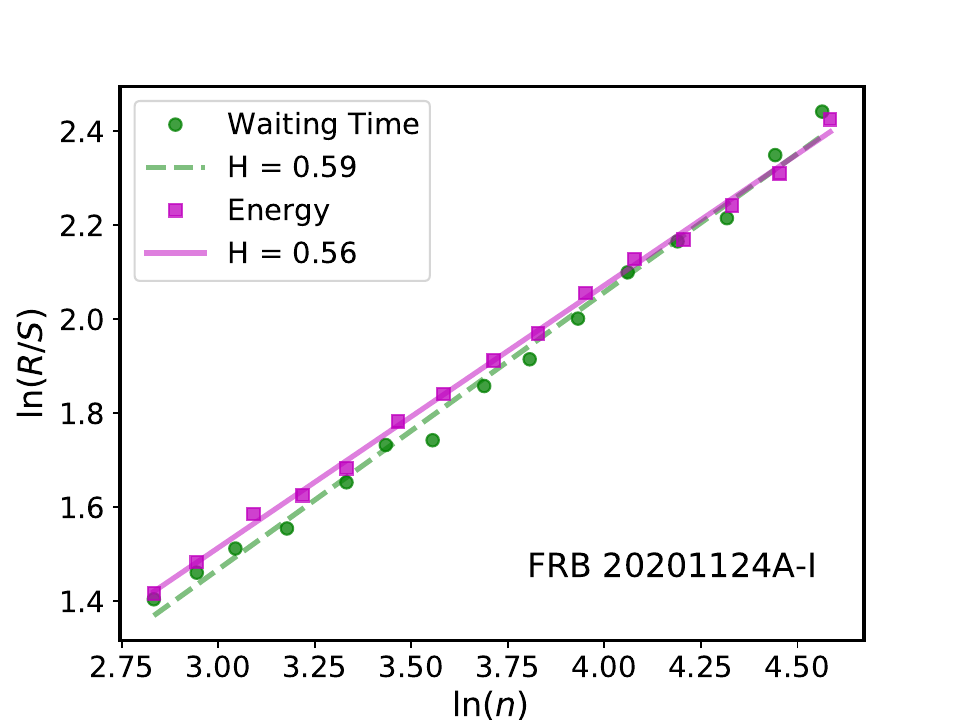}
    \includegraphics[width=0.48\textwidth]{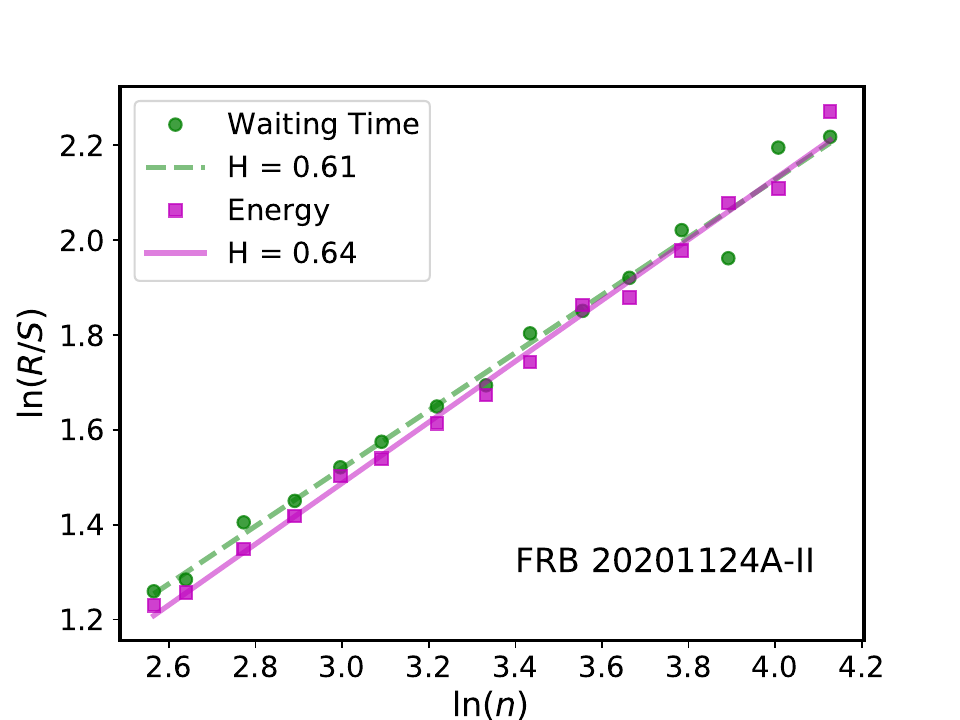}
    \includegraphics[width=0.48\textwidth]{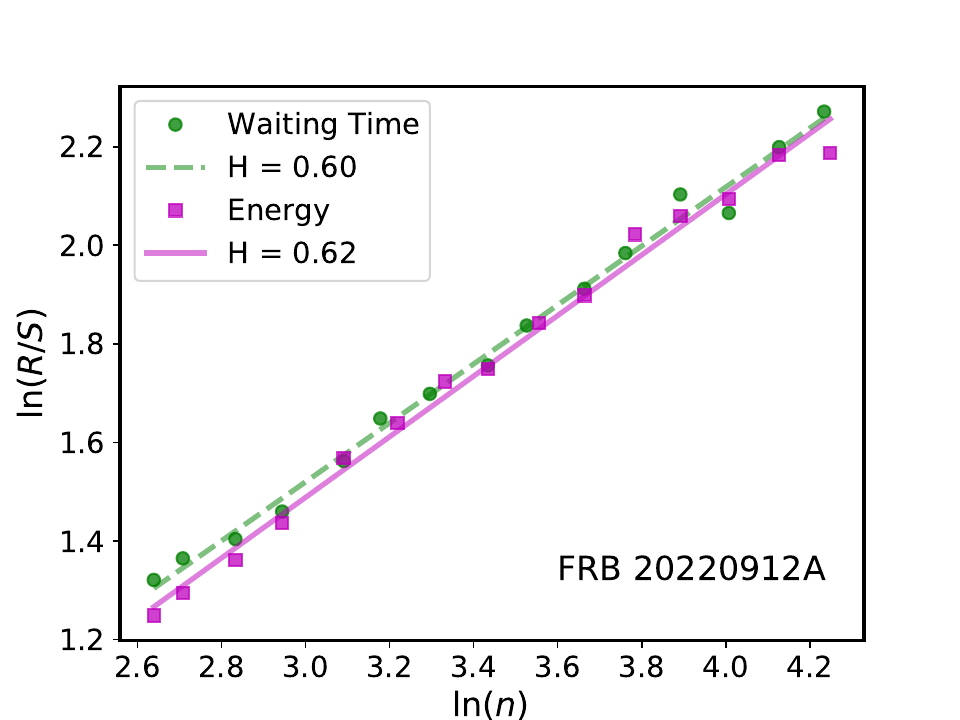}
    \caption{The rescaled range series for waiting time (green dots) and energy (magenta squares) of the four repeating FRB samples. The green dashed lines and magenta solid lines are the linear regression curve of waiting time and energy, respectively.}
    \label{fig:hurst}
\end{figure*}

We use the public code \textsf{NOLDS} \citep{scholzel_2020_3814723} for the calculation of Hurst exponent. The rescaled range series are shown in Figure \ref{fig:hurst}. The green dots and magenta squares are the rescaled range series of waiting time and energy, respectively. The green dashed lines and magenta solid lines are the corresponding linear regression curves. As can be seen, the data points are well consistent with the linear regression curves. The Hurst exponent $H$ is the slope of the best-fitting line, which is derived using the least squares method. Table \ref{tab:hurst} summaries the Hurst exponent $H$ for waiting time and energy of the four repeating FRB samples. The uncertainty represents the intrinsic scatter $\sigma_\mathrm{int}$, defined as the square root of the reduced chi-square. The Hurst exponents are $0.71 \pm 0.03$, $0.59 \pm 0.03$, $0.61 \pm 0.04$, $0.60 \pm 0.03$ for waiting time, and $0.63 \pm 0.02$, $0.56 \pm 0.01$, $0.64 \pm 0.03$, $0.62 \pm 0.03$ for energy for the samples of FRB 20121102A, FRB 20201124A-I, FRB 20201124A-II, and FRB 20220912A, respectively. For all the four repeating FRB samples, the Hurst exponents of waiting time and energy are larger than 0.5 at $\gtrsim 3\sigma$ confidence level, indicating the exist of long-term memory. However, the $H$ values are around 0.6, and do not deviate too much from 0.5, hence the long-range correlation is positive but not strong.

\begin{table}
\centering
\caption{The Hurst exponent $H$ for waiting time and energy of the four repeating FRB samples. The error is the intrinsic scatter $\sigma_\mathrm{int}$.}
\label{tab:hurst}
\begin{tabular}{lllll} 
\hline\hline 
 & Waiting time  & Energy \\
\hline
FRB 20121102A & 0.71 $\pm$ 0.03 & 0.63 $\pm$ 0.02\\
FRB 20201124A-I & 0.59 $\pm$ 0.03 & 0.56 $\pm$ 0.01 \\
FRB 20201124A-II & 0.61 $\pm$ 0.04 & 0.64 $\pm$ 0.03 \\
FRB 20220912A & 0.60 $\pm$ 0.03  & 0.62 $\pm$ 0.03 \\
\hline
\end{tabular}
\end{table}

\section{Randomness and Pincus Index}\label{sec:pincus}

The Pincus index \citep{Pincus1991App} is used to quantify the randomness of a data series without prior information about the source of dataset generation. The Pincus index is defined by the ratio of the approximate entropy of the original data sequence, to that of the randomly shuffled data sequence. For a data sequence ${u(i)}$ of length $N$, define the subsequences $x_{m}(i)=[u(i), ..., u(i + m - 1)]$ and $x_{m}(j)=[u(j), ..., u(j+m-1)]$. The approximate entropy is calculated as \citep{Pincus1991App,e21060541}
\begin{align}
   \nonumber
    ApEn(m,r,N) \simeq & -\frac{1}{N-m} \\ 
   &\times \sum_{i=1}^{N-m}\log \frac{\sum_{j=1}^{N-m}\{{\rm dist}[x_{m+1}(j), x_{m+1}(i)]<r\}}{\sum_{j=1}^{N-m} \{{\rm dist}[x_{m}(j), x_{m}(i)]<r\}}. 
\end{align}
Here ${\rm dist}[x,y]$ is the Chebyshev distance between $x$ and $y$, $\{\mathrm{condition}\}$ is 1 if the condition is true and 0 otherwise, and ``log" is the natural logarithm. The embedding dimension $m$ is the length of subsequences, usually taken to be 2 in the calculation. The radius distance threshold $r$ is the noise filter. The calculation of $ApEn$ for FRBs in this paper is based on the open-source toolkit \textsf{EntropyHub} \citep{EntropyHub}. To avoid the arbitrariness of the selection of $r$ value, the maximum approximate entropy ($MAE$) was used \citep{delgado2019quantifying,e21060541}, which quantify the largest difference in the information between the sequences of length $m$ and $m+1$. $MAE$ is given by varying the noise filter $r$ and calculating the maximum value of approximate entropy,
\begin{equation}
    MAE = \max_r  [ApEn(m,r,N) ].
\end{equation}
We follow a traditional approach to determine the maximum. Specifically, we vary $r$ from 0.01$\sigma$ to 0.99$\sigma$ with step 0.01$\sigma$, where $\sigma$ is the standard deviation of the data sequence, and choose the $r$ value which can maximize $ApEn(m,r,N)$.

To compare the randomness of different data sequences, we use bootstrap sampling and calculate the Pincus Index  \citep{Pincus1991App,delgado2019quantifying}. The original sequence are randomly shuffled for 1000 times and $MAE$ is calculated for each of the shuffled sequence. The Pincus index is given by
\begin{equation}
    PI = \frac{MAE_\textrm{original}}{ MAE_\textrm{shuffled}}.
\end{equation}
The $PI$ value measures the distance between $MAE_\textrm{original}$ and the histogram of $MAE_\textrm{shuffled}$. A completely ordered system gives a $PI$ value of zero, while the value is one or greater than one for a totally random system.

\begin{figure*}
    \centering
	\includegraphics[width=0.48\textwidth]{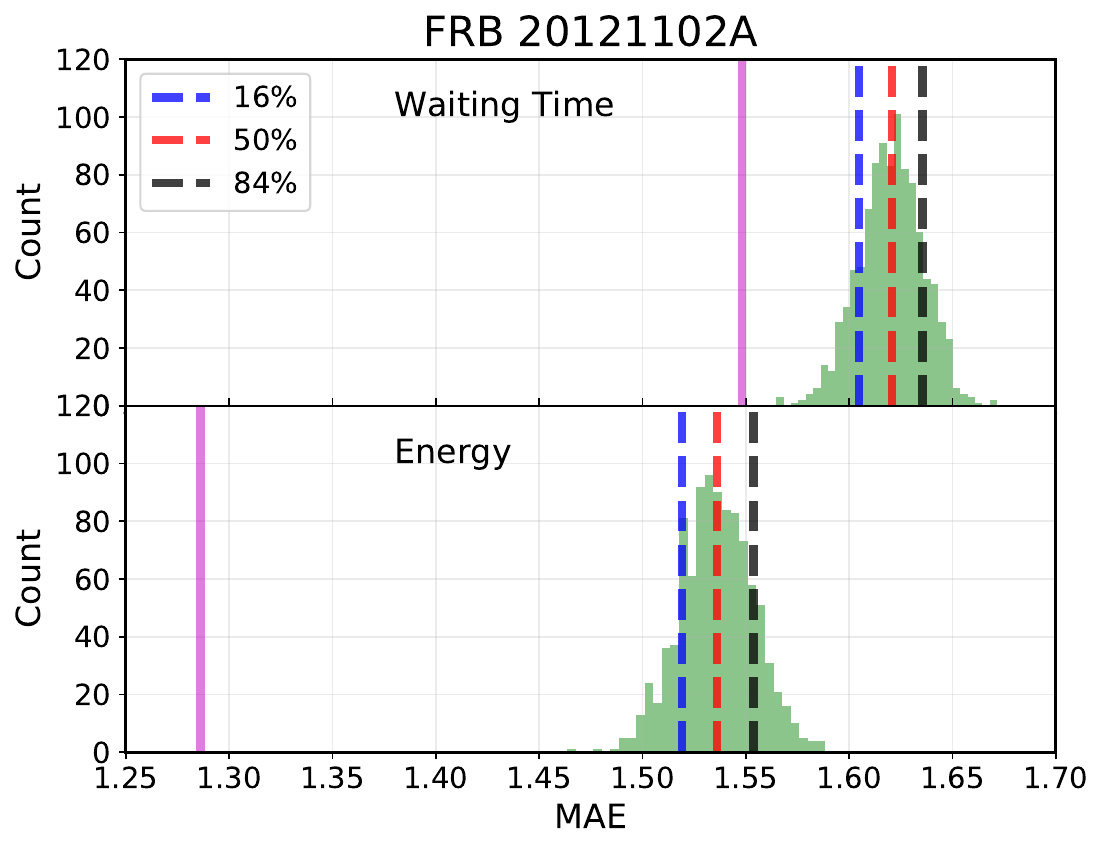}
	\includegraphics[width=0.48\textwidth]{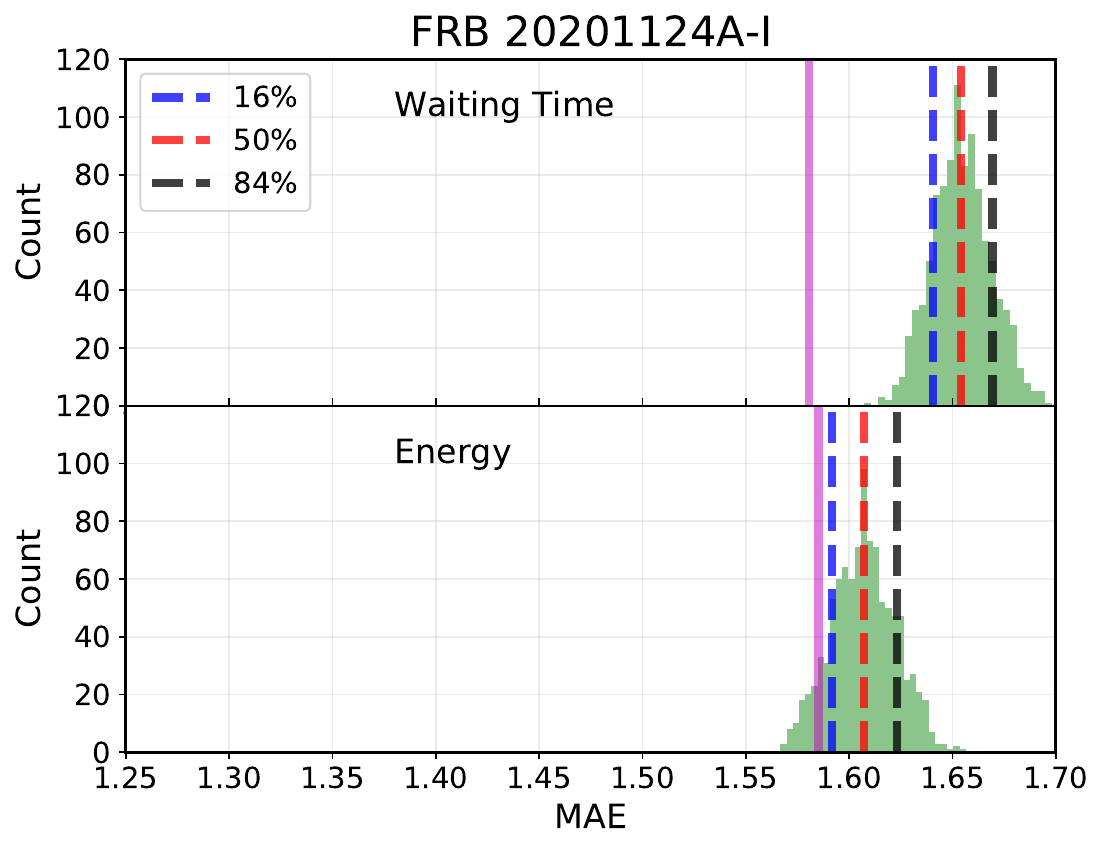}
    \includegraphics[width=0.48\textwidth]{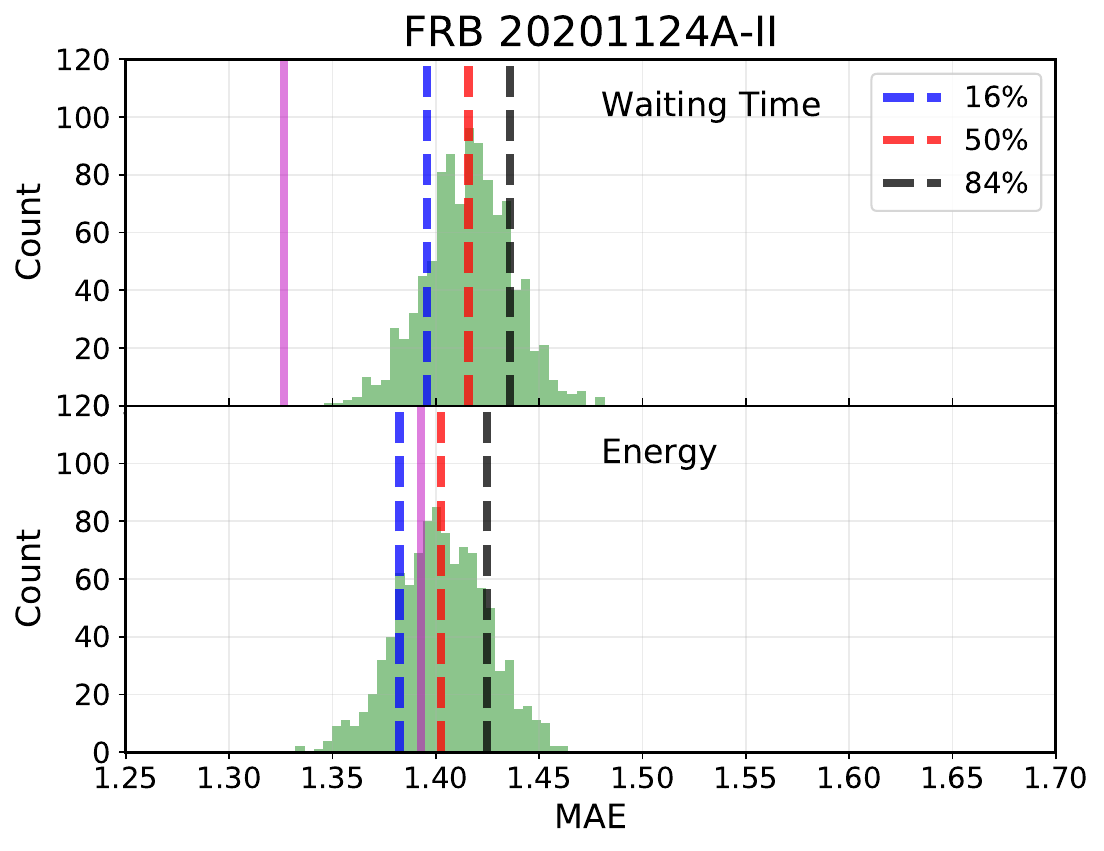}
    \includegraphics[width=0.48\textwidth]{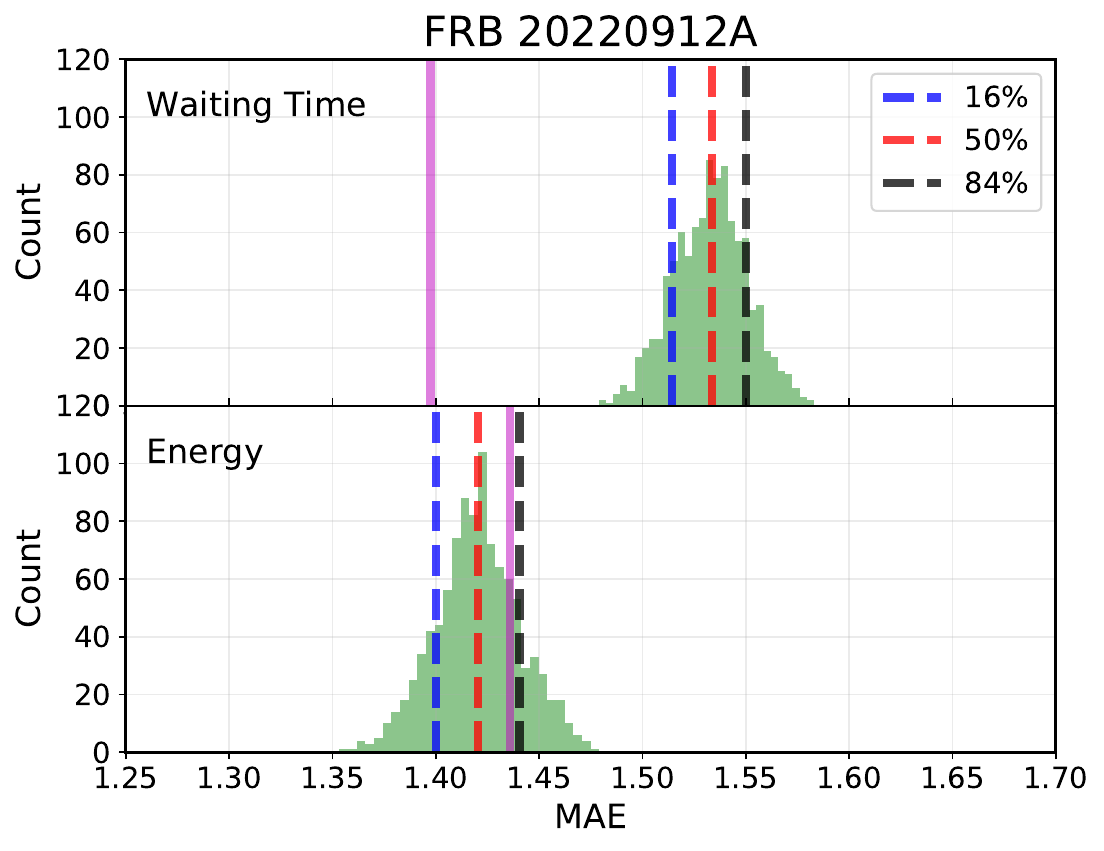}
    \caption{The $MAE$ distribution for the four repeating FRB samples. The green histogram is the distribution of $MAE$ of the 1000 shuffled sequences. The magenta solid line is $MAE$ of the original sequence. The blue, red and black dashed lines are the 16th, 50th and 84th percentile of the 1000 simulations, respectively.}
    \label{fig:pi}
\end{figure*}

Figure \ref{fig:pi} shows the $MAE$ distribution for the four repeating FRB samples.
The magenta solid line is the $MAE$ of the original sequence, and the green histogram is the $MAE$ of the 1000 randomly shuffled sequences. The blue, red and black dashed lines are the 16th, 50th and 84th percentile of the 1000 simulations, respectively. 
Table \ref{tab:pi} summaries the $MAE$ values of the original sequence and the $PI$ values for waiting time and energy. The uncertainty of $PI$ is propagated from the uncertainty of $MAE$, namely
\begin{equation}
\frac{\sigma_{PI}}{PI} = \frac{\sigma_{MAE_\textrm{shuffled}}}{ MAE_\textrm{shuffled}},
\end{equation}
where $\sigma_{MAE_\textrm{shuffled}}$ is the standard deviation of $MAE$ of the 1000 shuffled sequences.

\begin{table}
\centering
\caption{The $MAE$ of original sequence and $PI$ for waiting time and energy of  the four repeating FRB samples. The error of $PI$ is propagated from the standard deviation of $MAE$ of the 1000 randomly shuffled sequences.}
\label{tab:pi}
\begin{tabular}{lcccc} 
\hline\hline 
\multirow{2}*{ } & \multicolumn{2}{c}{Waiting time} &\multicolumn{2}{c}{Energy}\\ 
\cline{2-3}\cline{4-5}
& $\mathrm{MAE}_\mathrm{ori}$ &  PI & $\mathrm{MAE}_\mathrm{ori}$ & PI   \\
\hline
FRB 20121102A & 1.55 & 0.96 $\pm$ 0.01 & 1.29 & 0.84 $\pm$ 0.01 \\
FRB 20201124A-I & 1.58 & 0.96 $\pm$ 0.01 & 1.59 & 0.99 $\pm$ 0.01 \\
FRB 20201124A-II & 1.33 & 0.94 $\pm$ 0.01 & 1.40 & 0.99 $\pm$ 0.02 \\
FRB 20220912A & 1.40  & 0.91 $\pm$  0.01 & 1.44 & 1.01 $\pm$ 0.01 \\
\hline
\end{tabular}
\end{table}

As shown in Figure \ref{fig:pi}, it is easy to quantity the randomness by comparing the $MAE_\textrm{original}$ and $MAE_\textrm{shuffled}$. For example, the energy of FRB 20201124A-II is consistent with a totally random organization because the value of $MAE_\textrm{original}$ is very close to the 50th percentile of the $MAE_\textrm{shuffled}$ histogram. Similarly, the energies of FRB 20201124A-I and FRB 20220912A are also consistent with random organization. However, the energy of FRB 20121102A is significantly different from a random organization, because the distance from the histogram is considerable. These conclusions can also be seen from the $PI$ value, as is summarized in Table \ref{tab:pi}. The energy of FRB 20121102A with $PI=0.84$ is less random than the energy of the other three FRBs with $PI\approx 1.0$. As for the waiting time, although the $PI$ values of all the four samples are close to 1.0, they deviate from 1.0 at more than $3\sigma$ confidence level. This implying that the waiting time of all the four FRB samples are not completely random.

\citet{Li:2021hpl} found that the bimodal log-normal waiting time distribution of FRB 20121102A can be reproduced by randomly generated samples, which suggested the randomly emission of bursts. However, the simulation is not completely random but takes into account the observational effects. The time of arrivals were randomly simulated through Monte Carlo method, following the exact setup of the observations, e.g., the starting time, duration, sampling rate, and pulse burst rate. In our paper, we conclude that the waiting times of the four FRB samples are not completely random. Our calculation is based on the $PI$ value, which quantifies the randomness of a data series without prior information about the source of dataset generation. Moreover, \citet{Li:2021hpl} didn't consider the temporal order of the waiting time. Randomly shuffling a data sample does NOT change the probability distribution, but DO change the $PI$ value. For example, if a data sample $X=\{x_1,x_2,\cdots,x_n\}$ is randomly drawn from e.g. Gaussian distribution, we may say that $X$ is random. If we sort the same data sample $X$ in a monotonically increasing order, then the sorted $X$ still follows the Gaussian distribution, but it is obviously not random. Although the bimodal waiting time distribution of FRB 20121102A can be reproduced by a random process, it does not necessarily mean that the data sample is completely random.

\section{Self-organized criticality and $q$-Gauss of fluctuation}\label{sec:fluctuation}

The fluctuations of energy and waiting time provide important information about the self-organized criticality (SOC) \citep{Bak:1987xua,Bak:1989Earth,aschwanden2011self}. The probability density function of fluctuations of, e.g. energy and waiting time in a SOC system, is expected to follow the Tsallis $q$-Gaussian function but not the Gaussian function, and the $q$ values are invariant for different temporal intervals \citep{Caruso:2007PhRvE}. It has already been shown that the fluctuations of energy and waiting time of FRB 20121102A is scale invariant \citep{Lin:2019ldn}, consistent with the characteristics of SOC. In this section, we further investigate the SOC property of repeating FRBs using the updated samples.

\begin{figure*}
    \centering
    \includegraphics[width=0.48\linewidth]{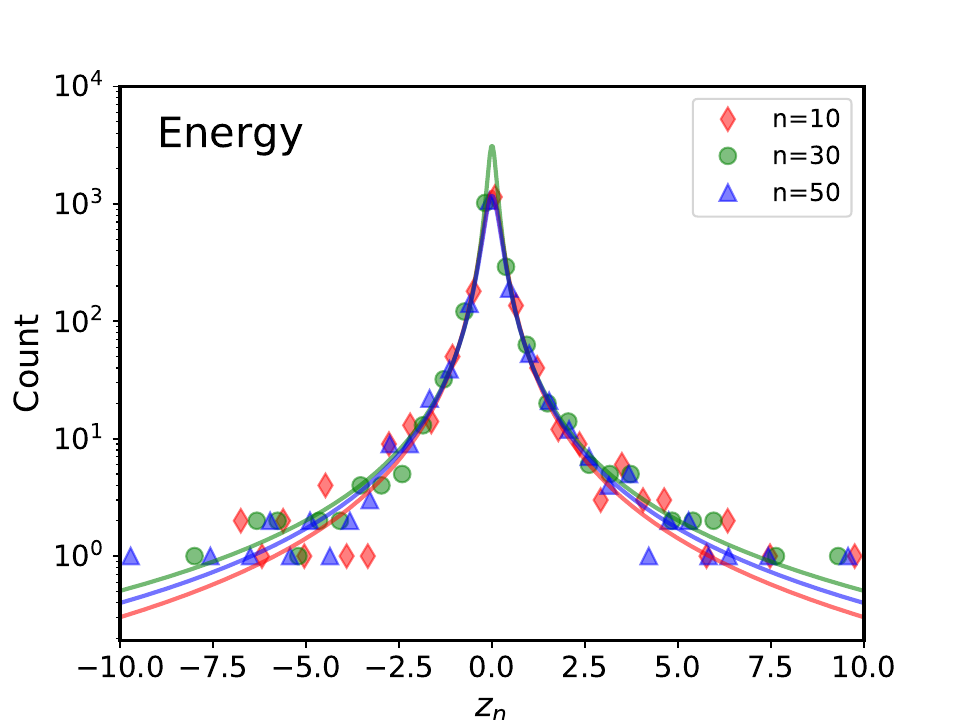}
    \includegraphics[width=0.48\linewidth]{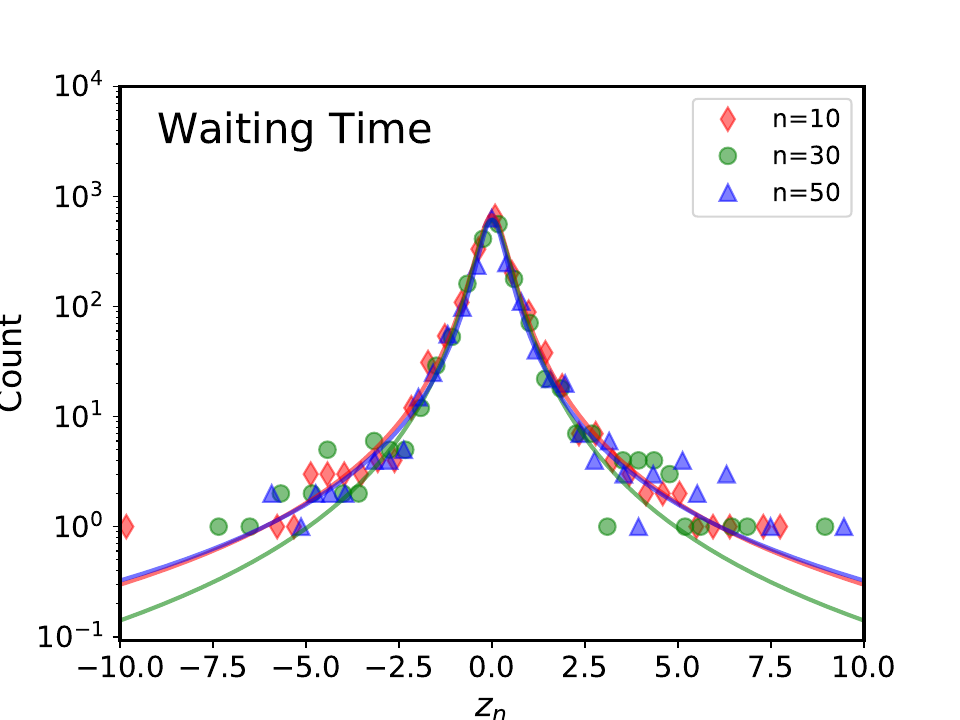}
    \caption{The $q$-Gaussian distributions of fluctuations of energy (left panel) and waiting time (right panel) for FRB 20121102A.}
    \label{fig:fluctuation_121102}
\end{figure*}

The fluctuation of a time series $\{Q_i\}$ ($i=1,2,\cdots,N)$ is defined by $\{Z_n=Q_{i+n}-Q_i\}$, where $n=1,2,\cdots$ is the temporal interval scale. In order to compare quantities of different magnitudes and dimensions, the fluctuation is usually normalized by its standard deviation, defining the dimensionless fluctuation $z_n=Z_n/{\rm std}(Z_n)$. Figure \ref{fig:fluctuation_121102} shows the fluctuations of energy and waiting time for FRB 20121102A, with three different temporal interval scales $n=10,30,50$. We can see that the distribution of fluctuation shows a sharp peak near $z_n=0$, and have a flat tail at both left and right sides. These features are distinctly different from the Gaussian distribution.

The distribution can be well fitted with the Tsallis $q$-Gaussian function \citep{Tsallis:1987eu,Tsallis:1998ws},
\begin{equation}
  f(x)=\alpha[1-\beta(1-q)x^2]_+^{\frac{1}{1-q}},
\end{equation}
where $\alpha$, $\beta$ and $q$ are free parameters, and $[x]_+\equiv{\rm max}(0,x)$. The parameters $q$ and $\beta$ control the shape and width of the peak, respectively, and $\alpha$ is a normalization constant. The $q$-Gaussian distribution is a generalization of the standard Gaussian distribution, and in the limit $q\rightarrow 1$ it reduces to the Gaussian distribution with mean value $\bar{x}=0$ and standard deviation $\sigma=1/\sqrt{2\beta}$. For $q<1$ the distribution is bounded in a finite range, while for $q>1$ it extends to infinity. If $q\geq5/3$, the second moment and the variance do not exist. The $q$-Gaussian distribution is widely used in various fields, such as physics, astronomy, geology, economics, and so on.

The solid lines in Figure \ref{fig:fluctuation_121102} show the $q$-Gaussian fits to the data points. As can be seen, the $q$-Gaussian distribution can well mimic the sharp peak and flat tails of the data points. The data points extends to a large value and no obvious boundary can be seen, so the best-fitting $q$-value is greater than 1.

\begin{figure*}
    \centering
    \includegraphics[width=0.48\linewidth]{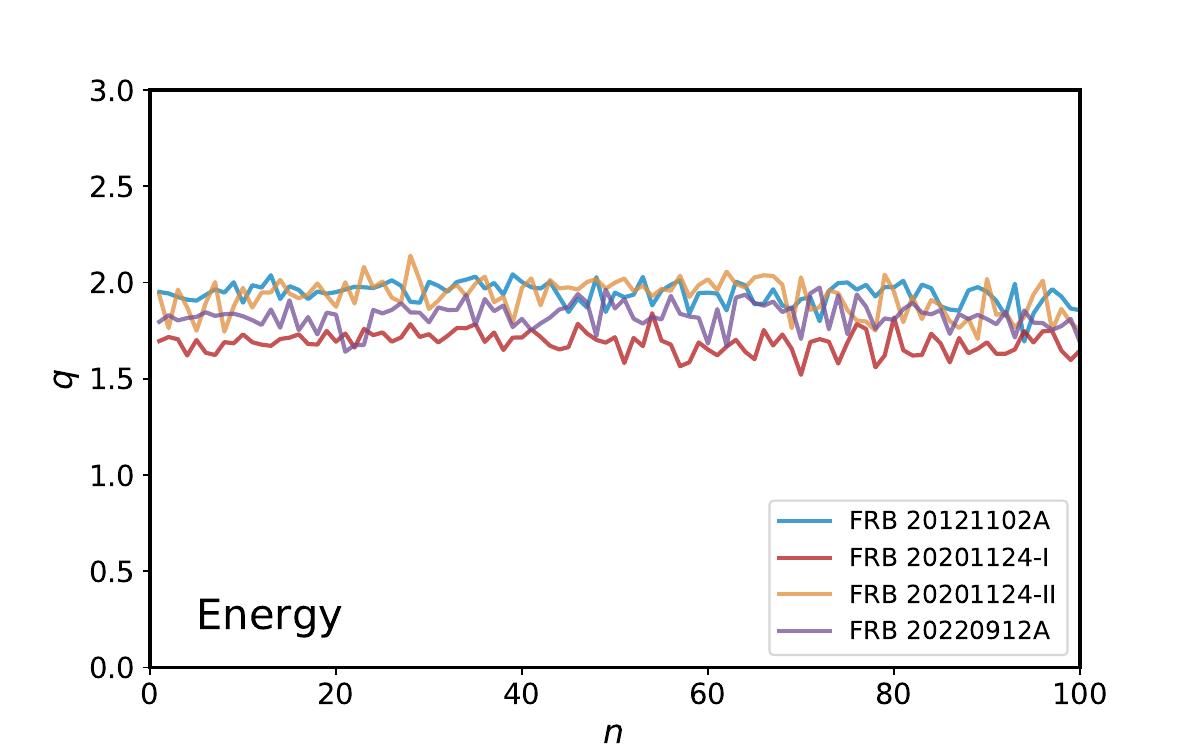}
    \includegraphics[width=0.48\linewidth]{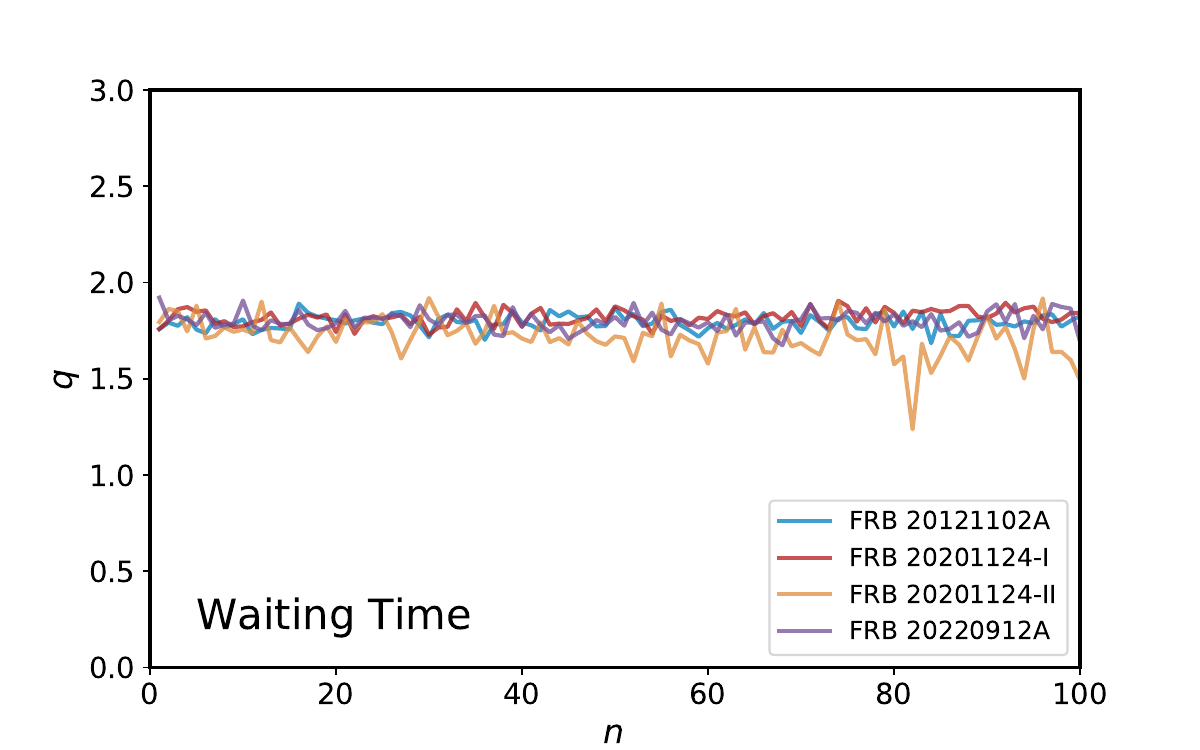}
    \caption{The best fitting q-value for the fluctuation of energy (left panel) and waiting time (right panel) as a function $n$.}
    \label{fig:qvalues}
\end{figure*}

The distribution of fluctuation depends on the temporal interval scale $n$. In Figure \ref{fig:qvalues}, we plot the best fitting $q$-value for the fluctuations of energy (left panel) and waiting time (right panel) as a function $n$. The results of four FRBs are plotted together for comparison. We can see that the $q$-value approximately keeps steady when $n$ varies, implying that the fluctuation is scale-invariant. The average $q$ values of the $q$-Gaussian distribution are summarized in Table \ref{tab:qvalues}. For the fluctuation of energy, the average $q$-values of all the four samples are consistent with each other, falling into a smaller range  $q\approx 1.7\sim 1.9$. For the fluctuation of waiting time, the average $q$-values of all the four samples are also consistent, with $q\approx 1.7\sim 1.8$, although FRB 20201124A-II shows a relatively high variance at large $n$.

\begin{table}
    \centering
    \caption{The average $q$ values of the $q$-Gauss distribution. The uncertainties represent the standard deviations.}
     \label{tab:qvalues}
    \begin{tabular}{lll}
    \hline\hline
         & Energy & Waiting time\\
     \hline
     FRB 20121102A    & $1.94\pm0.06$ & $1.79\pm0.04$ \\
     FRB 20201124A-I  & $1.69\pm0.06$ & $1.82\pm0.04$ \\
     FRB 20201124A-II & $1.92\pm0.09$ & $1.72\pm0.10$ \\
     FRB 20220912A    & $1.83\pm0.07$ & $1.80\pm0.05$ \\
     \hline
     \end{tabular}
\end{table}

\section{Discussions and Conclusions}\label{sec:conclusion}
In this paper, we investigated the statistical properties of FRBs using four data samples from three extremely active repeating FRBs observed by the FAST telescope. Three statistics are used to quantify the randomness of the FRB activity. 
1) The rescaled range analysis method was used to calculate the Hurst exponent of energy and waiting time. We found the existence of long-term memory in the time series of energy and waiting time in all the four data samples. 
2) We calculated the maximum approximate entropy and Pincus index to quantify the randomness of waiting time and energy. We found that the waiting time slightly deviates from the random organization for all the four samples. A significant deviation from the random organization was found in the time series of energy for FRB 20121102A. However the energy for the other three samples are consistent with a totally random organization.
3) The fluctuations of energy and waiting time of all the four samples are well described by the Tsallis $q$-Gaussian distribution. The $q$ value is independent of the temporal intervals, consistent with the scale-invariance property of the self-organized criticality.

The existence of long-term memory of repeating FRBs has already been cross-checked by different methods. 
Based on the coherent growths in burst-rate structures, \citet{Wang:2023wcb} found that two repeating FRBs (FRB 20121102A and FRB 20201124A) reveal memory over timescales from a few minutes to about an hour. In a recent work, \citet{Wang:2023sjs} studied the conditional mean waiting time and the conditional mean residual time to the next burst and reported the memory with a temporal clustering. The above two works studied the memory in the occurrence time, i.e., the time series of waiting time. In this paper, we found that the long-term memory exits not only in waiting time, but also in burst energy.
The long-term memory was also found in earthquakes \citep{barani2018long}. For example, it was shown that seismicity is a memory process with a Hurst exponent $H \approx 0.87$ based on the Hurst’s rescaled range analysis of series of cumulative seismic moment both in Italy and worldwide \citep{barani2018long}. The Hurst exponent of earthquakes ($H \approx 0.87$) is larger than the repeating FRBs discussed in our paper, indicating a more pronounced long-term correlation in earthquakes.

In a recent work by \citet{Zhang:2023fmn}, the randomness in the time series of energy was compared between repeating FRBs, pulsars, solar flares, earthquakes, and Brownian motion. It was found that the randomness of FRB 20190520B ($PI=0.97$) is similar to the Brownian motion ($PI=0.99$), but FRB 20121102A ($PI=0.84$) is evidently different from the Brownian motion. The conclusion of \citet{Zhang:2023fmn} is consistent with the results of our calculations, e.g., see the $PI$ values of energy listed in Table \ref{tab:pi}. Except for FRB 20121102A, the randomness of energy of the rest three samples (FRB 20201124A-I, FRB 20201124A-II and FRB 20220912A) is well consistent with Brownian motion. This implies that FRB 20121102A is peculiar compared with other repeating FRBs, but the physical reason is still unclear. In the time domain, the PI values of the four samples are all larger than $0.9$, implying a large degree of randomness in burst activity. But considering the small uncertainty, the PI values of all the four bursts deviate from 1.0 at more than $3\sigma$ confidence level, indicating that the burst activity of repeating FRBs is not completely analogue to Brownian Motion. The statistics of the two data sample from the same source (FRB 20201124A-I and FRB 20201124A-II) are consistent with each other, implying no obvious temporal evolution of burst activity, at least in this two observation periods.

The randomness of magnetar bursts has also been investigated based on the $PI$ values both for time and energy domains \citep{Yamasaki:2023fud}. In the energy domain, the energy fluctuation instead of the energy itself was used to calculate the PI values, which is different from the work of \citet{Zhang:2023fmn} and our paper. It was found that randomness of energy fluctuation of magnetar bursts exhibits a broad consistency with FRBs. In the time domain, \citet{Yamasaki:2023fud} calculate the $PI$ values of waiting time for magnetar bursts, as what were done for repeating FRBs in the work of \citet{Zhang:2023fmn} and our paper. The $PI$ values of waiting time are 0.53, 0.52 and 0.47 for three magnetar samples, respectively, which are smaller than the $PI$ values ($PI \sim 0.91-0.96$) of waiting time for FRBs obtained in our paper, indicating a significantly lower degree of randomness in the time domain for magnetar bursts.

The fluctuations of energy and waiting time of FRBs follows the $q$-Gaussian distribution. For all the four samples, the $q$ values keep steady over temporal scales, falling into the range $q\approx 1.7\sim 1.9$. The scale invariance of fluctuation is a significant feature of the SOC system, which has already been widely investigated for some geological and astronomical events, such as earthquakes \citep{Caruso:2007PhRvE,Wang:2015nsl}, soft-gamma repeaters (SGRs) \citep{Chang:2017bnb,Sang:2021cjq,Wei:2021kdw}, and repeating FRBs \citep{Lin:2019ldn,Wang:2022gmu}. For instance, \citet{Chang:2017bnb} found that the $q$ value of energy (fluence) fluctuation of SGR J1550–5418 is around $q\approx 2.4$. Similar results are found in SGR 1935+2154, with the $q$ value in the range $q\approx 2.2\sim 2.3$ \citep{Sang:2021cjq}. \citet{Caruso:2007PhRvE} found that the $q$ value of energy fluctuation of earthquake is around $q\approx 1.75$, which is well consistent with the $q$ value of FRBs we found here. Using two small data samples from FRB 20121102A observed by different telescopes, \citet{Lin:2019ldn} found that the $q$ value of energy fluctuation is around $q\approx 2.0$, which is a little larger than the values of the four FRB samples we investigate here.

In summary, all the three statistics show that the apparently irregular repeating FRBs are actually not completely random. There exits long-term memory in the FRB activity. The scale invirance of fluctuation of energy and waiting time is consistent with the feature of the SOC systems. One of the most interesting models displaying SOC is the Olami-Feder-Christensen (OFC) model \citep{olami1992self}. The scale invariant $q$-Gaussian distribution of fluctuations can be explained by the OFC model on a small world topology \citep{Caruso:2007PhRvE}. We expect that the future observations on FRBs will be helpful to provide new clues to the occurrence mechanism of repeating FRBs.

\section*{Acknowledgements}

This work has been supported by the National Natural Science Fund of China (Grant Nos. 12005184, 12175192, 12275034 and 12347101), and the Fundamental Research Funds for the Central Universities of China under grant no. 2024CDJXY-022.

%%%%%%%%%%%%%%%%%%%%%%%%%%%%%%%%%%%%%%%%%%%%%%%%%%
\section*{Data Availability}

The data of FRB 20121102A is available in \citet{Li:2021hpl}, the data of FRB 20201124A-I is available in \citet{Xu:2021qdn}, the data of FRB 20201124A-II is available in \citet{Zhang:2022rib}, and the data of FRB 20220912A is available in \citet{Zhang:2023eui}.



\bibliographystyle{mnras}
\bibliography{reference} % if your bibtex file is called reference.bib

%\begin{thebibliography}{99}
%\end{thebibliography}

%%%%%%%%%%%%%%%%%%%%%%%%%%%%%%%%%%%%%%%%%%%%%%%%%%

%%%%%%%%%%%%%%%%% APPENDICES %%%%%%%%%%%%%%%%%%%%%
%\appendix

%\section{extra figures}\label{sec:appendix}

%%%%%%%%%%%%%%%%%%%%%%%%%%%%%%%%%%%%%%%%%%%%%%%%%%

% Don't change these lines
\bsp	% typesetting comment
\label{lastpage}
\end{document}